\documentclass{IEEEtran}
\renewcommand{\thesubsection}{\Alph{subsection}\itshape}
\renewcommand{\thesubsubsection}{\arabic{subsubsection}}

\usepackage{spconf,amsmath,graphicx,comment}
\usepackage{float}

\title{Probing Mental Health Information in Speech Foundation Models}
\name{\begin{tabular}{c} Marc de Gennes$^{\star}$ \qquad Adrien Lesage$^{\star}$ \qquad Martin Denais \qquad Xuan-Nga Cao \\ Simon Chang \qquad Pierre Van Remoortere \quad Cyrille Dakhlia \quad Rachid Riad\thanks{$\star$ Equal contribution.} \end{tabular}}
\address{\begin{tabular}{c}
Callyope \\
\small\texttt{\{marc, adrien, rachid\}@callyope.com}
\end{tabular}}

\begin{document}
%\ninept
%
\maketitle
\begin{abstract}
 Non-invasive methods for diagnosing mental health conditions, such as speech analysis, offer promising potential in modern medicine. Recent advancements in machine learning, particularly speech foundation models, have shown significant promise in detecting mental health states by capturing diverse features. This study investigates which pretext tasks in these models best transfer to mental health detection and examines how different model layers encode features relevant to mental health condition. We also probed the optimal length of audio segments and the best pooling strategies to improve detection accuracy. Using the Callyope-GP and Androids datasets, we evaluated the models' effectiveness across different languages and speech tasks, aiming to enhance the generalizability  of speech-based mental health diagnostics. Our approach achieved SOTA scores in depression detection on the Androids dataset.

\end{abstract}
\begin{keywords}
Probing, Foundation model, Mental health, Depression, Pooling
\end{keywords}

\label{sec:intro}

% These guidelines include complete descriptions of the fonts, spacing, and
% related information for producing your proceedings manuscripts. Please follow
% them and if you have any questions, direct them to Conference Management
% Services, Inc.: Phone +1-979-846-6800 or email
% to \\\texttt{papers@2024.ieeeicassp.org}.

\section{Introduction}
%rr: 
%lintroduction est trop longue et verbeuse, il faudrait eviter les commentaires 'traditional models'  : OK

%est ce que vous pouriez articuler lenchainement de cette intro en plan ou phrases cles qui senchainent? quest ce que chaque paragraphe apporte ? 

%le probleme nest pas assez explicitee ? what and how mental health information is encoded in foundation speech models ? 

%dans la partie dataset n= au milieu du texte cest bizarre ca ne se fait pas

%'diving deeper in the model' cest etrange comme formulation aussi

Major Depressive Disorder (MDD) affects between 30 and 40\% of individuals at some point in their lifetime, across sexes and worldwide \cite{gbd2022global}. It is linked to lower global functioning, physical comorbidities and premature death. 
Non-invasive methods in mental healthcare  offer potential for better clinical decision-making and improved patient care \cite{guo2015measurement}. In particular, speech analysis represents a powerful avenue for identifying mental health conditions \cite{cummins2015review, williamson2013vocal, cohn2009detecting} in a non-invasive fashion. The human voice carries rich information about a speaker's psychological state \cite{weintraub2023word}. Predefined acoustic markers have been crafted based on the knowledge of the speech production system  \cite{eyben2015geneva} and have been proven to be useful for depression detection \cite{ringeval2019avec}. Technological advances with deep learning models for instance, now enable the automatic discovery of more complex speech patterns, further improving the accuracy of mental health detection \cite{ringeval2019avec, deng2017speech, tasnim2024machine}.  

  Foundation models like HuBERT \cite{hsu2021hubert}, wav2vec \cite{baevski2020wav2vec}, and Whisper \cite{radford2023robust} are pre-trained on large, diverse datasets with various pretext tasks. Therefore, they can easily adapt  to multiple tasks, even in low-data regime. They have proven to be highly effective at extracting detailed acoustic and linguistic features \cite{bommasani2021opportunities}. These models not only excel at encoding speaker-specificity information but can also generalize tasks by effectively capturing broader linguistic and speech patterns \cite{gong2023whisper}. In particular, recent studies have shown promising results in identifying subtle speech patterns linked to mental health conditions \cite{wu2023self, gerczuk2023zero}.

In this study, we probed how (and which) foundation models manage to assess individuals' mental state. We examined models across all layers to explore which specific representations could capture features indicative of mental health condition. We found that, to the best of our knowledge, these models are comparable or surpass the state of the art (SOTA). Additionally, we probed the amount of information these models require to operate: how much speech is necessary to detect mental health states? Is information present in every spoken utterance or only in specific segments? Does lexical content contribute for more accurate detection? To address these questions, we examined the length of audio context that is needed to capture meaningful mental health indicators. We then focused on optimal pooling strategies to combine information across multiple audio segments, and used the mellowmax function \cite{pmlr-v70-asadi17a} to explore a continuous spectrum of strategies between the traditional min, mean, and max pooling selection. We tested the role of semantics by using both elicited and spontaneous production tasks. Finally, we evaluated the generalizability of our findings by comparing two datasets: the French-speaking Callyope-GP Corpus, which contains audio recordings from the general population, and the Italian Androids Corpus, featuring recordings from clinically depressed patients \cite{tao2023androids}. This approach allowed us to assess mental health detection across different languages and populations.

The paper is structured as follows: We first introduce the pre-trained speech models selected for our experiments. Next, we describe the datasets and their relevance to our research objectives. Following this, we detail our probing methodology and experimental setup. Finally, we present our results and discuss their implications for advancing speech-based mental health applications.

%- main issue, depression detection, multilingual lack of studies, 
%- probing information from large  model
%x-vector probing \cite{raj2019probing} found: spoken content and channel-relate such as session of recording

%- probing of whisper for noise robustness, and found transfer very well to acoustic events, \cite{gong2023whisper} -> found large scale whisper and 

%- probing language, gender, phoneme representation in self supervised models \cite{de2022probing}

%- closest work: manifestation of depression overlaps with speaker representation \cite{dumpala2023manifestation} 

%- data sparsity angle + closest work but small  dataset waic oz with only students speech (here extended and extensive study) \cite{wu2023self}

\section{Pre-Trained Speech representations}
\label{sec:speech_models}

The  \begin{it}wav2vec2\end{it} \cite{baevski2020wav2vec} and \begin{it}HuBERT\end{it} \cite{hsu2021hubert} speech models both share a similar self-supervised training method: it relies predicting masked parts of the audio input in order to force the model to learn high level representations. The training sets are \begin{it}Librispeech\end{it} \cite{panayotov2015librispeech} and \begin{it}Libri-light\end{it} \cite{kahn2020libri}, which contain clean speech data from audio books. On the other hand, the \begin{it}Whisper\end{it} \cite{radford2023robust} speech model is trained to predict transcripts on a large dataset containing audio files collected on the internet. In this paper, we only used the encoder part of Whisper.

We evaluated the following versions of the forementioned models: wav2vec2 (wav2vec2-large-xlsr-53), HuBERT-L (hubert-large-ls960-ft) and HuBERT-XL (hubert-xlarge-ls960-ft), Whisper-M (whisper-medium) and Whisper-L (whisper-large-v3).

\section{Datasets}

In our study, we used two datasets: our own \textit{Callyope-GP Corpus} and the publicly available \textit{Androids Corpus} \cite{tao2023androids}. 

The \textit{Callyope-GP Corpus} comprises data from 860 French-speaking participants from the general population, who completed a series of speech tasks and self-administered questionnaires through the Callyope research mobile application. Two types of speech tasks were required: an elicited production task (also referred to as contained-vocabulary (CV) task), where participants were asked to count from 1 to 20, and a spontaneous speech task (also referred to as open-vocabulary (OV) task), where the following prompt was presented: “Describe how you are feeling at the moment and how your sleep has been lately”. The total duration of the dataset is 10.12 hours long, sampled at 16kHz.
The Callyope-GP dataset also includes self-administered questionnaires on the participant's mental health status, including depressive symptoms, anxiety, and insomnia. Depressive symptoms were measured using the Beck Depression Inventory (BDI) and the Patient Health Questionnaire-9 (PHQ-9), with a score of 10 or above serving as the threshold for classification in our predictions. Anxiety and insomnia levels were assessed with the Generalized Anxiety Disorder questionnaire (GAD-7) and the Athens Insomnia Scale (AIS), using  classification thresholds of 10 and 6, respectively. %Finally, overall fatigue was evaluated through the Multidimensional Fatigue Inventory (MFI), with a threshold score of 40. 
The dataset was randomly split into a training and a test set, and Table \ref{tab:demographics} summarizes its characteristics.
\begin{table}[h!]
\centering
\begin{tabular}{lcc}
\hline
\textbf{Clinical Evaluation} & \textbf{Training} & \textbf{Test} \\ \hline
Number of participants   &  731                 &     129          \\ \hline
$\text{BDI}>10$   &  228 (31.19\%) & 44 (34.11\%) \\ \hline
$\text{PHQ-9}>10$ & 122 (16.69\%)  & 23 (17.83\%) \\ \hline
$\text{GAD-7}>10$ & 115 (15.73\%)  & 16 (12.40\%) \\ \hline
$\text{AIS}>6$    &  314 (42.95\%) & 54 (41.86\%) \\ \hline
\end{tabular}
\caption{Characteristics of the Callyope-GP dataset.}
\label{tab:demographics}
\end{table}

The \textit{Androids Corpus} serves as a benchmark for automatic depression detection using speech \cite{tao2023androids}. It contains 228 recordings from 118 native Italian speakers, including 64 clinically diagnosed with depression by professional psychiatrists, hence providing a more accurate characterization of depressed patients compared to self-reported measures. The corpus includes both elicited and spontaneous speech samples, with a total duration of approximately 9 hours of data. The original paper uses a 5-fold cross-validation method, allowing for robust evaluation across multiple subsets of the data, and we kept those same splits in our experiments. 

\section{Probing experiments}
\label{sec:probing}

We investigated the most effective way to retrieve mental health information from audio representations. We assume that if this type of information is encoded in the embeddings, a simple linear classifier will correctly classify a target related to mental health \cite{raj2019probing}. We used a logistic regression as linear classifier, implemented through the sci-kit learn package \cite{pedregosa2011scikit}. The metric used is the average $F_1$ score of the classification task over the labels available for each corpus.

\subsection{Probing Embedder Representations}
\label{probing embedders}

The encoders of Whisper, HuBERT, and wav2vec2  store multiple intermediary representations that can be used as embeddings. The first probing experiment focuses on assessing which of these representations contain the most information about mental health condition. 

Each audio recording of the  two datasets is partitioned with a window size of 20s and an overlap of 10s between two consecutive windows. A prediction is computed for each window, and the prediction for the full audio is defined as the average of the predictions over all windows.

\subsection{Probing Temporal Dynamics}

The second probing experiment aims to determine the optimal context to capture mental health information from speech. Instead of giving a full audio to a speech model, the audio recording is commonly split into windows, and each window is in turn given to the model. This windowing method creates two hyper-parameters in the model.

The first parameter controls the length of the window and the overlap between two consecutive windows. In this paper, the overlap used is half the length of the window.

The second parameter is the method used to pool the different predictions made by the classifier on each of the windows. In addition to the traditional pooling methods $\operatorname{max}$, $\operatorname{min}$ and $\operatorname{mean}$, we used the mellowmax function that gives a continuous interpolation between min and max functions. The mellowmax of a vector $x$ is defined as: 
\begin{equation*}
    \operatorname{mm}_\omega(x)=\frac{1}{\omega}\log \left(\frac{1}{n} \sum_{i=1}^n e^{\omega x_i}\right).
\end{equation*}
The mellowmax converges to the maximum function when $\omega$ goes to infinity, to the minimum function when $\omega$ goes to minus infinity and to the mean function when it goes to zero.

We explored the links between these two hyper-parameters: context and ways of pooling. We conducted this experiment only on the spontaneous speech task, using the last layer of the HuBERT-XL and Whisper-L encoders. We opted for these two models because they have been trained with different methods, are widely used, and these particular versions generally achieve the highest scores. We settled on window sizes of 0.5s, 1s, 2s, 5s, 10s, 15s, and 20s and the pooling methods $\operatorname{max}$, $\operatorname{min}$, $\operatorname{mean}$ and $\operatorname{mm}_\omega$, for $\omega$ equals to $\pm 0.1$, $\pm 1$, $\pm 10$ and $\pm 100$.

\begin{comment}

\subsection{Probing duration}

Our third objective was to determine for each model the minimal duration of audio segment that contains information about mental health condition. 

Similarly to the second experiment, audio recordings werewindowed. Instead of pooling the predictions made after slicing the audios and computing the embeddings, we (arbitrarily) selected the prediction made on the middle window as the prediction for the full audio segment. 

We used windows sizes of 0.5s, 1s, 2s, 5s, 10s, 15s, 20s, 25s and 30s. We use the embedders wav2vec2 $\mathrm{LARGE}$, HuBERT $\mathrm{X-LARGE}$ and Whisper $\mathrm{LARGE}$.
\end{comment}

\section{Results and discussion}
\label{sec:results}

\subsection{Probing Embedder Representations}

\begin{figure}
    \centering
    \includegraphics[width=1.0\linewidth]{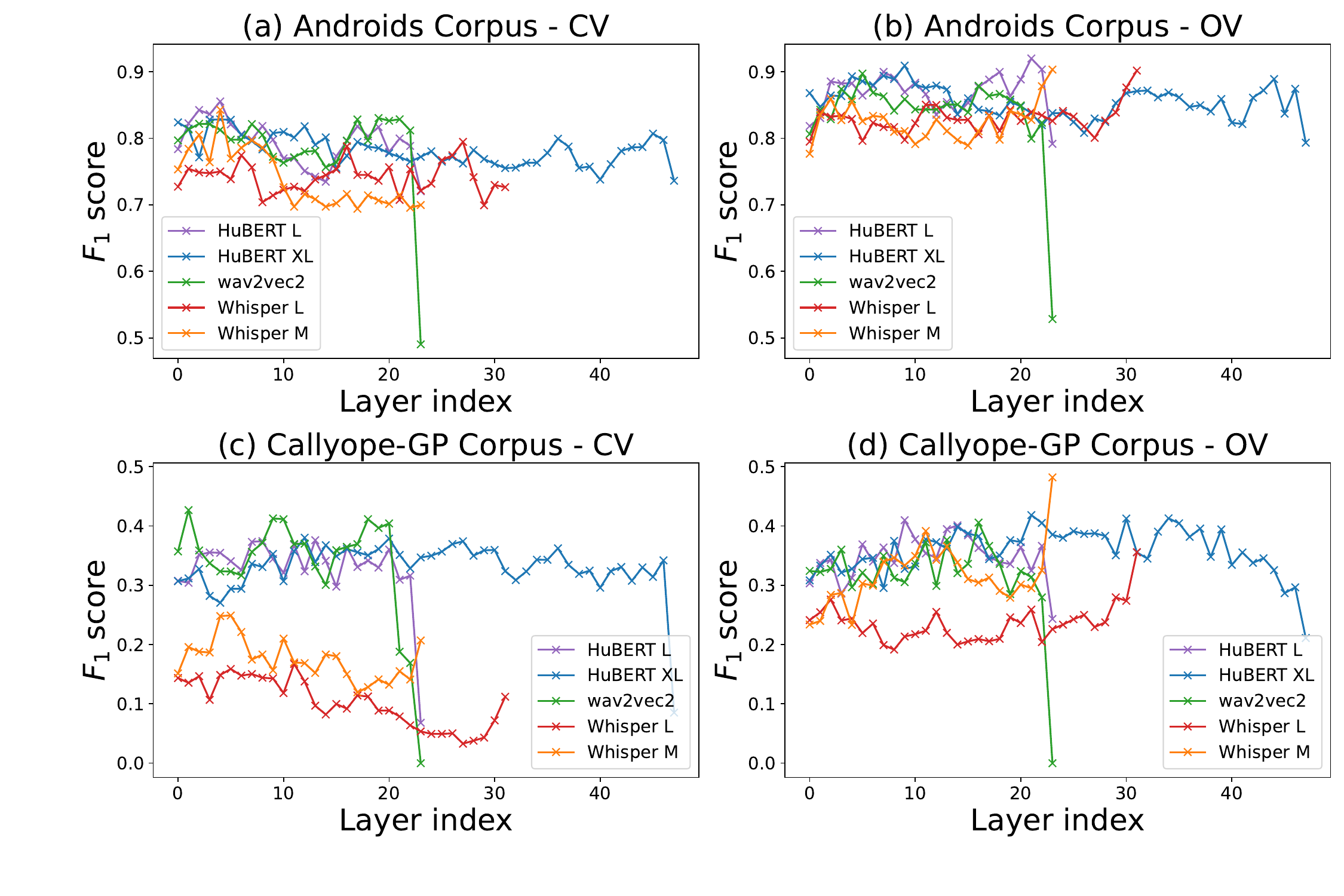}
        \caption{Performance of encoders across their layers. Each graph shows the mean $F_1$ score as a function of the index layer for different models. Graphs (a) and (b) use the Androids corpus, while (c) and (d) use the Callyope-GP corpus. Graphs (a) and (c) correspond to the spontaneous task, and (b) and (d) correspond to the elicited task.}
    \label{fig:first_experiment}
\end{figure}

The results are displayed in Figure \ref{fig:first_experiment}. 

For the Androids corpus (Fig.~1(a) and 1(b)), performance did not vary much among the different layers of each encoder, except for wav2vec's last layer. HuBERT-L had the best performance on the elicited production task (Fig.~1(a), $F_1=0.86$), which is comparable to the SOTA \cite{tao2023relationship}. It also performed best on the spontaneous production task with the layer 21 (Fig.~1(b), $F_1=0.92$), which, to the best of our knowledge, surpasses the current SOTA \cite{tao2023androids} (Table \ref{tab:model_performance}). wav2vec2's last layer scored the worst on both tasks. The average performance across layers and models is better on the spontaneous task.

For the Callyope-GP corpus (Fig.~1(c) and 1(d)), performance did not vary much for all models, except for the last layer of these three models: HuBERT-L, HuBERT-XL and wav2vec2. Both versions of the Whisper model perform significantly worse than the other models on the elicited production task (Fig.~1(c)). However, Whisper-M performed best on the spontaneous production task (Fig.~1(d)), thanks to its last layer, which is significantly higher than its preceding layers. Wav2vec2 and both HuBERT models displayed the lowest score with their last layer on both tasks.

Poor performance on the last layer of wav2vec2,  compared to the other layers, could be explained by the fact that it was specifically added after  pre-training and fine-tuned to perform a specific task of speech recognition \cite{baevski2020wav2vec}.

Both versions of Whisper performed better on the spontaneous task for each corpus. It is particularly distinguishable in the Callyope-GP corpus (Fig.~1(c) and 1(d)), where Whisper-M earned the best performance on the spontaneous task. On that same task, Whisper-L also achieved its best scores thanks to the last layers of its encoder, but this strong performance was not replicated for the elicited production task. This may be because Whisper, trained primarily for speech recognition, encodes more semantic information, which is irrelevant in elicited speech tasks. It is also noteworthy that on spontaneous task, Whisper models reach their higher performance on their last layers, which is closer to the decoder and might contain the more semantic information.

Overall, on both tasks and for both corpora, we observed results that are not as polished as in \cite{gong2023whisper}, where the authors performed a similar layer probing experiment with sound classification. For each model, the variation in performance from one layer to the next is more erratic.

\subsection{Probing Temporal Dynamics}
\label{subsec:temporal}

\begin{figure}
    \centering
    \includegraphics[width=1.0\linewidth]{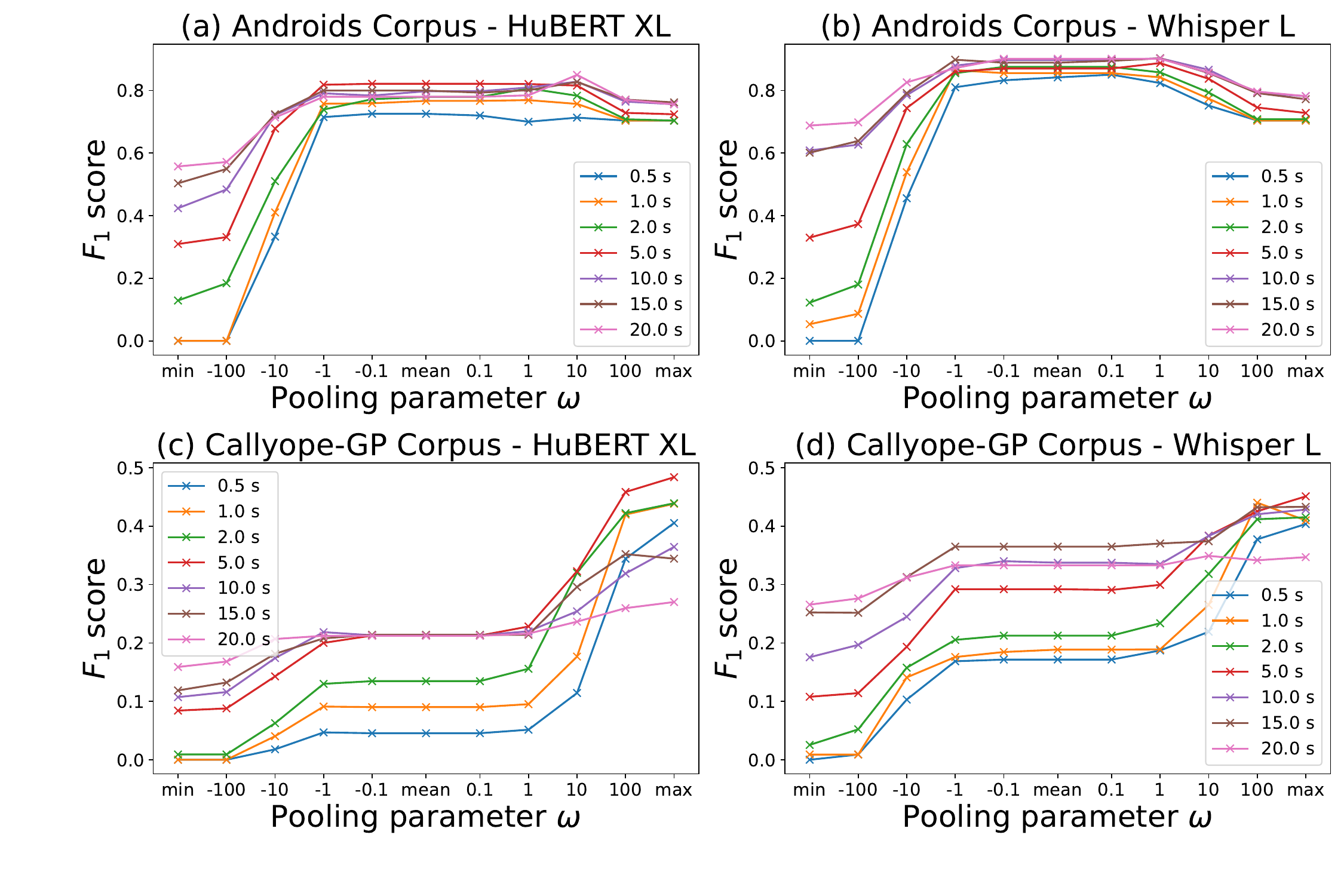}
    \caption{Effect of window size and pooling on performance. Each graph shows the mean $F_1$ score as a function of the pooling parameter $\omega$, for  window sizes ranging from 0.5s to 20s. Graphs (a) and (b) use the Androids corpus, while (c) and (d) use the Callyope-GP corpus. Graphs (a) and (c) correspond to the HuBERT-XL model, and (b) and (d) correspond to the Whisper-L.}
    \label{fig:second_experiment}
\end{figure}

\begin{table*}[h!]
\centering
\begin{tabular}{lcccccccccccc}
\hline
\textbf{Corpus} & \multicolumn{10}{c}{\textbf{Callyope-GP}} & \multicolumn{2}{c}{\textbf{Androids}} \\ \hline
                \textbf{Measure} & \multicolumn{2}{c}{\textbf{BDI}} & \multicolumn{2}{c}{\textbf{PHQ-9}} & \multicolumn{2}{c}{\textbf{GAD-7}}  & \multicolumn{2}{c}{\textbf{AIS}} & \multicolumn{2}{c}{\textbf{Mean}} & \multicolumn{2}{c}{\textbf{Diagnosis}} \\ \hline
                \textbf{Audio task} & \textbf{CV} & \textbf{OV} & \textbf{CV} & \textbf{OV} & \textbf{CV} & \textbf{OV} & \textbf{CV} & \textbf{OV} & \textbf{CV} & \textbf{OV}  & \textbf{CV} & \textbf{OV} \\ \hline
HuBERT L & 0.49 & 0.59 & 0.42 & 0.3 & 0.29 & 0.36 & 0.50 & \textbf{0.65} & 0.38 & 0.41 & 0.86 & \textbf{0.92} \\ \hline
HuBERT XL & 0.48 & 0.55 & 0.43 & \textbf{0.36} & 0.26 & 0.38 & \textbf{0.60} & 0.61 & 0.38 & 0.42 & 0.83 & 0.91 \\ \hline
wav2vec2 & \textbf{0.51} & 0.5 & \textbf{0.49} & 0.26 & \textbf{0.33} & 0.39 & 0.55 & 0.64 & \textbf{0.43} & 0.41 & 0.83 & 0.90 \\ \hline
Whisper M & 0.29 & \textbf{0.60} & 0.26 & 0.33 & 0.12 & \textbf{0.40} & 0.46 & 0.60 & 0.25 & \textbf{0.48} & 0.84 & 0.90 \\ \hline
Whisper L & 0.22 & 0.46 & 0.16 & 0.15 & 0.0 & 0.19 & 0.37 & 0.62 & 0.17 & 0.36 & 0.79 & 0.90 \\ \hline
Tao et al.\cite{tao2023relationship} & &&&&&&&&&& \textbf{0.88}
 \\ \hline
Tao et al.\cite{tao2023androids} & &&&&&&&&&&& 0.85 \\ \hline
\end{tabular}
\caption{Model performance across the Callyope-GP and Androids datasets, measured by $F_1$ score. "Mean" refers to the average of BDI, PHQ-9, GAD-7, and AIS scores. OV and CV denote open-vocabulary and constrained-vocabulary audio tasks. The best performing layer is selected for each measure, with hyperparameters as in section \ref{sec:probing}.\ref{probing embedders}}
\label{tab:model_performance}
\end{table*}

%\subsubsection{Optimal window size}

When screening for the optimal window size, we observed that short windows of 0.5 second consistently performed poorly, across all conditions (Fig. \ref{fig:second_experiment}, blue curves). This suggests that a very short context did not provide enough information to capture meaningful mental health indicators.
For the Callyope-GP corpus, the optimal window size was 5 seconds (Fig. \ref{fig:second_experiment} (c,d), red curves), yielding the best $F_1$ scores for both HuBERT-XL ($F_1=0.48$) and Whisper-L ($F_1=0.45$). In contrast, using the longest window size of 20 seconds resulted in poorer performance, with $F_1 = 0.24$ for HuBERT-XL and $F_1=0.33$ for Whisper-L.\\
On the Androids corpus, the models achieved good performance over a broader range of audio lengths (Fig. \ref{fig:second_experiment}(a,b)). Both HuBERT-XL and Whisper-L reached their top scores with a window size of 20 seconds ($F_1=0.85$ and $F_1=0.90$), though similar scores were also obtained across the 5s - 15s range. 

The probing of pooling strategies showed that results were largely independent of the model used.

For the Androids corpus, both mean and max poolings yielded strong performance. HuBERT-XL reached its top performance for a value $\omega=10$ of the pooling parameter (Fig. \ref{fig:second_experiment}(a), pink curve), corresponding to an intermediate of mean and max pooling. Whisper-L had its best performances for mean pooling (Fig. \ref{fig:second_experiment}(b), $\omega=0$). 

In contrast, for the Callyope-GP corpus (Fig. \ref{fig:second_experiment}(c,d)), the $F_1$ score was an increasing function of $\omega$, and best scores were always obtained with the max pooling strategy.  

Two potential reasons could explain these differences in pooling performance across datasets. First, the Callyope-GP dataset is imbalanced, with the negative class over-represented for each questionnaire (see Table \ref{tab:demographics}). In imbalanced scenarios, models tend to over-predict the majority class (negative cases), and max pooling helps capturing rare instances where a few audio slices exhibit strong positive signals. We tested this hypothesis by replicating our experiment on a balanced Callyope-GP dataset, created by undersampling the negative class: the performance gap between mean and max pooling was indeed reduced (see Figure \ref{fig_appendix} in the Appendix).

Second, participants from the Callyope-GP corpus come from the general population, where even positive cases likely show milder symptoms compared to the Androids dataset, which includes patients diagnosed and monitored by clinicians. Therefore, the "positive" signal in Callyope-GP may be weaker and more sporadic, favoring max pooling, which captures such sparse signals more effectively. This second aspect may also explain why longer window sizes of 20 seconds, which average a lot of information, performed poorly on the Callyope-GP dataset.

\section{Conclusion}

This study explored the capabilities of speech foundation models in detecting mental health conditions, by probing their layers and evaluating different pooling strategies.
Our findings demonstrate that these models can effectively capture information relevant to mental health, with performances comparable to or surpassing the current SOTA. All models showed better performance on spontaneous audio tasks, suggesting that semantics may play a key role in detection. Whisper, specifically trained for speech recognition, performed significantly better on these spontaneous tasks. Its last layers, being closest to the text decoder, are particularly rich in semantic content, and showed the best performance across all layers.

Our experiments underscore the importance of tailoring model configurations to the characteristics of the dataset. For more severe patients, as in the Androids corpus, longer audio windows and mean pooling yielded better results, while for the imbalanced Callyope-GP dataset, where patients likely exhibit milder symptoms, max pooling and shorter audio windows proved more effective.

One limitation of our work is the potential effect of multilingual factors, as our datasets represent different languages (French and Italian). It remains unclear if and how language differences contributed to the variation in results across corpora.

Future work could involve fine-tuning foundation models specifically for mental health detection tasks, aiming to improve their accuracy and generalizability across various populations, mental health conditions and languages.

\vfill\pagebreak
\small

\bibliographystyle{IEEEbib}
\bibliography{strings}

\appendix 

\section{Appendix}
In this appendix, we summarize our extended analysis of speech foundation models applied to the Callyope-GP and Androids datasets, to provide a broader picture concerning window sizes, pooling and balance of data. 
\begin{figure}[h!] \centering \includegraphics[width=0.8\linewidth]{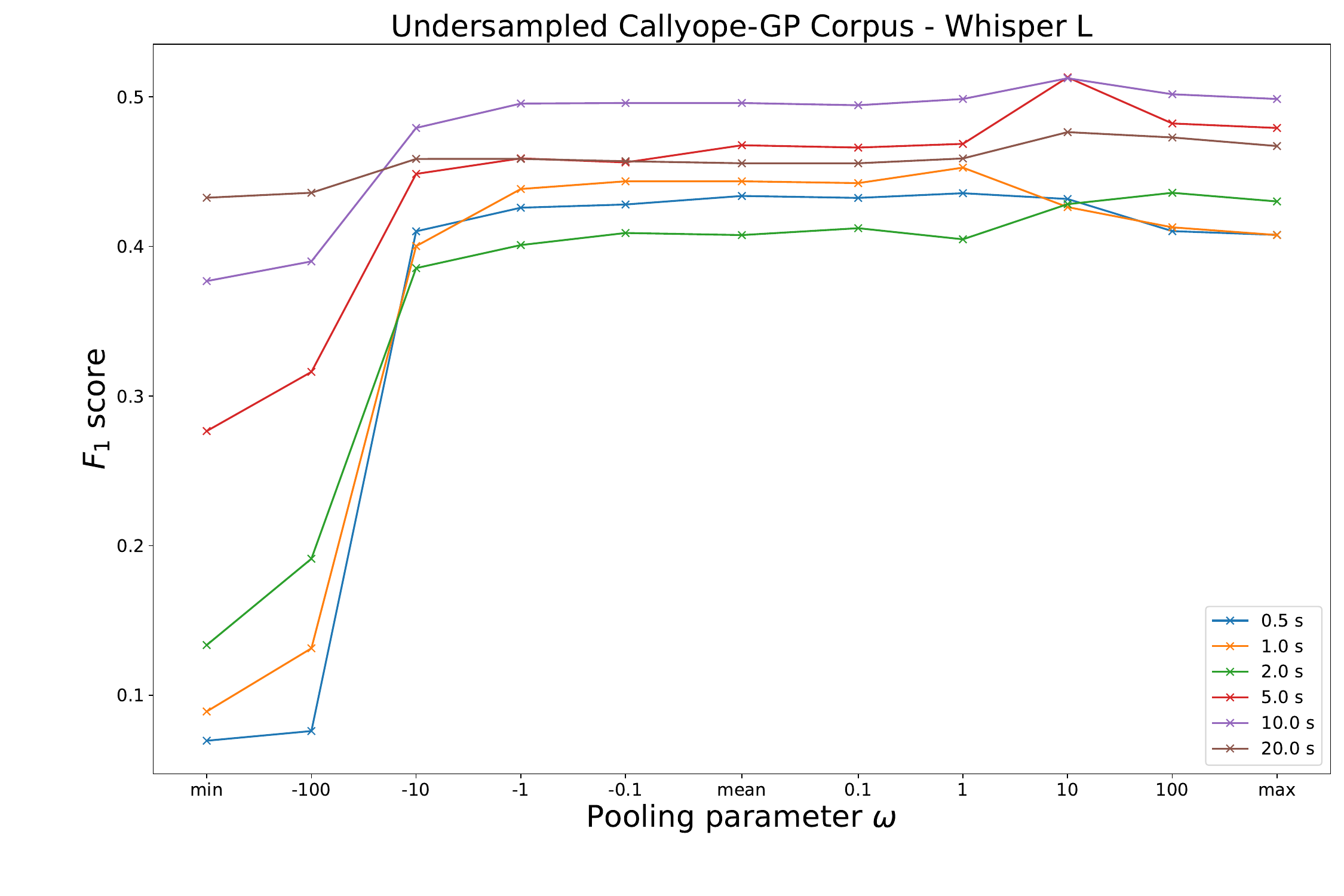} \caption{Effect of window size and pooling on a balanced version of the Callyope-GP Corpus. The datasets are balanced by randomly undersampling the majority class. Parameters are the same as in Figure \ref{fig:second_experiment}(d).}
\label{fig_appendix}
\end{figure}

\subsection{Effect of Pooling on Balanced Datasets} \label{appendix_1}

In Section \ref{sec:results}.{\ref{subsec:temporal}}, we hypothesized that the success of max pooling for the Callyope-GP corpus was linked to the imbalance of positive vs negative cases. Therefore, we tested the impact of dataset imbalance on pooling strategy. For each questionnaire of the Callyope-GP corpus (see Table \ref{tab:demographics}), we ensured equal representation of both classes by randomly undersampling the majority class. We only performed this experiment with Whisper-L, as pooling effects on the Callyope-GP corpus were found to be mostly model-independent (see Figure \ref{fig:second_experiment}(c,d)).

As shown in Figure \ref{fig_appendix}, undersampling significantly improves overall performance: a smaller but more balanced dataset yields better results (highest $F_1$ = 0.52 compared to 0.45 in Figure \ref{fig:second_experiment}(d)). Although this best performance is achieved for a mellowmax function close to max pooling ($\omega=10$, window size of 5 or 20 seconds), the gap between max and mean pooling is notably reduced. This confirms that class imbalance in the Callyope-GP corpus plays a critical role in determining the optimal pooling strategy. 

\subsection{Optimal Window Size without Pooling} 
\label{appendix_2}

In Section \ref{sec:results}.{\ref{subsec:temporal}}, we examined the combined effects of window size and pooling strategies. Here, we isolated the impact of window size by evaluating single-window segments, without any pooling. We focused on the central window of each audio recording, with window sizes ranging from 0.5 to 30 seconds. We analyzed the last layers of HuBERT-XL and Whisper-L, and the penultimate layer of wav2vec2.

\begin{figure}[H] 
\centering
\includegraphics[width=0.8\linewidth]{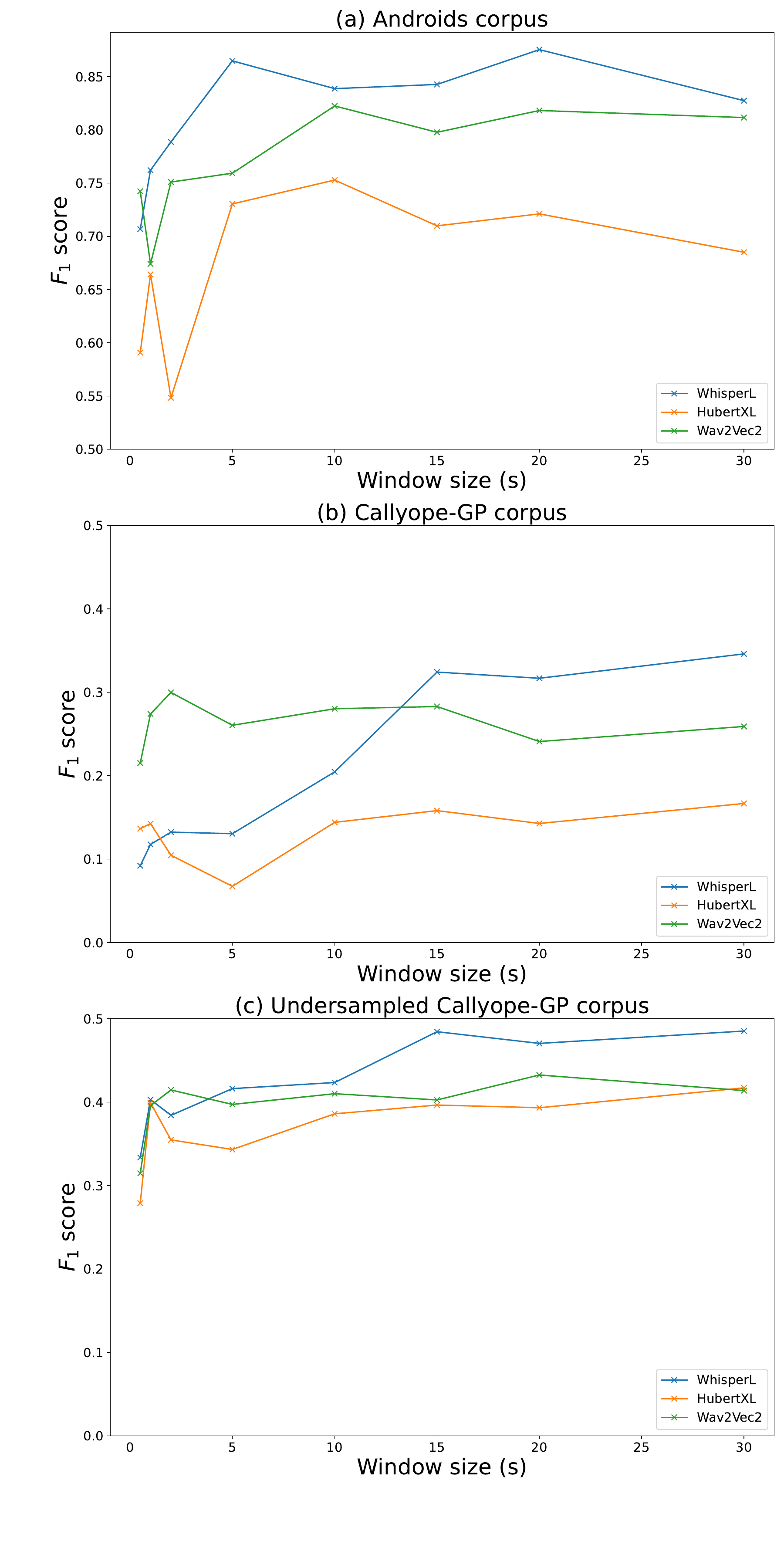} 
\caption{Probing the performance of models on a single window, for different window sizes. (a) On the Androids Corpus, models show high performance even with as little as 5 or 10 seconds of audio. (b) On the Callyope-GP corpus, performance is significantly reduced, highlighting the importance of pooling and sampling strategies. (c) Undersampling the Callyope-GP corpus significantly improves performance, with the best results obtained for longer audio window sizes.}
\label{fig:duration}
\end{figure}

On the Androids corpus, all models maintained high performance, even with limited context (Figure \ref{fig:duration}(a)). With only a 5-second window, Whisper-L achieved high performance ($F_1=0.86$), while HuBERT-XL and wav2vec2 reached their peak performance with 10-second windows ($F_1=0.75$ and $F_1=0.82$, respectively).

In contrast, the Callyope-GP corpus showed a significant performance drop using this approach, especially for HuBERT-XL (Figure \ref{fig:duration}(b)). This is consistent with the earlier findings where undersampling and pooling strategies greatly improved our results. We measured this effect of undersampling in Figure \ref{fig:duration}(c), which resulted in a significant improvement. For Whisper-L, predictions performed on a single segment in the middle of the audio yielded a score $F_1=0.49$, outscoring all pooling methods and window sizes explored on the imbalanced Callyope-GP corpus in Figure \ref{fig:second_experiment}. Interestingly, the best performances occured for larger window sizes —20 seconds for wav2vec2, and 30 seconds for Whisper-L and HuBERT-XL— which suggests that isolated segments of the callyope-GP dataset may lack sufficient information to fully capture mental health indicators.

\end{document}